\documentclass[conference]{IEEEtran}
\IEEEoverridecommandlockouts
\usepackage[utf8]{inputenc}
\usepackage{amsmath}
\usepackage{amsfonts}
\usepackage{amssymb}
\usepackage{mathtools}
\usepackage{amsthm}
\usepackage{fontenc}
\usepackage{graphicx}
\usepackage{balance}
\usepackage{booktabs}
\usepackage{subfigure}
\usepackage{tabulary}
\usepackage{booktabs}
\usepackage[justification=centering]{caption}
\usepackage{comment}
\usepackage{cite}
\usepackage{algorithmic}
\usepackage{algorithm}
\usepackage[algo2e]{algorithm2e} 
\usepackage[bookmarks=false]{hyperref}

\newtheorem{Prob}{\textbf{Problem}}

\newtheorem{Def}{Definition}
\SetKw{KwBy}{by}
\usepackage[colorinlistoftodos]{todonotes}
\usepackage{caption}
\usepackage[left=0.68in,right=0.68in,top=0.7in, bottom=1.0in]{geometry}

\usepackage[nodisplayskipstretch]{setspace}

\title{Soft Actor-Critic Learning-Based Joint Computing, Pushing, and Caching Framework in MEC Networks}

\author{Xiangyu Gao, Yaping Sun, Hao Chen, Xiaodong Xu, Shuguang Cui,~\IEEEmembership{Fellow,~IEEE}
\thanks{The work was supported in part by NSFC with Grant No. 62293482, the Basic Research Project No. HZQB-KCZYZ-2021067 of Hetao Shenzhen-HK S$\&$T Cooperation Zone, NSFC with Grant No. 62301471, The Major Key Project of PCL Department of Broadband Communiation, 
the National Key R$\&$D Program of China with grant No. 2018YFB1800800, the Shenzhen Outstanding Talents Training Fund 202002, the Guangdong Research Projects No. 2017ZT07X152 and No. 2019CX01X104, the Guangdong Provincial Key Laboratory of Future Networks of Intelligence (Grant No. 2022B1212010001), the Shenzhen Key Laboratory of Big Data and Artificial Intelligence (Grant No. ZDSYS201707251409055), and the Key Area R$\&$D Program of Guangdong Province with grant No. 2018B030338001.}

\thanks{X. Gao is with the University of Washington, Seattle, USA. (email: xygao@uw.edu) Y.~Sun and H.~Chen are with the Department of Broadband Communication, Peng Cheng Laboratory, Shenzhen 518000, China. (email: \{sunyp, chenh03\}@pcl.ac.cn) \textcolor{black}{Y.~Sun is also with the Future Network of Intelligent Institute (FNii)}, the Chinese University of Hong Kong (Shenzhen), Shenzhen 518172, China. X.~Xu and P.~Zhang are  with the Beijing University of Posts and Telecommunications, Beijing 100876, China, and affiliated with the Department of Broadband Communication, Peng Cheng Laboratory, Shenzhen 518000, China. (email: xuxiaodong, pzhang@bupt.edu.cn) S.~Cui is with the School of Science and Engineering (SSE) and \textcolor{black}{the Future Network of Intelligent Institute (FNii)}, the Chinese University of Hong Kong (Shenzhen), Shenzhen 518172, China. S.~Cui is also  affiliated with the Department of Broadband Communication, Peng Cheng Laboratory, Shenzhen 518000, China (email: shuguangcui@cuhk.edu.cn).
}}

\begin{document}
\maketitle

\begin{abstract}
To support future 6G mobile applications, the mobile edge computing (MEC) network needs to be jointly optimized for computing, pushing, and caching to reduce transmission load and computation cost. To achieve this, we propose a framework based on deep reinforcement learning that enables the dynamic orchestration of these three activities for the MEC network. The framework can implicitly predict user future requests using deep networks and push or cache the appropriate content to enhance performance. To address the curse of dimensionality resulting from considering three activities collectively, we adopt the soft actor-critic reinforcement learning in continuous space and design the action quantization and correction specifically to fit the discrete optimization problem. We conduct simulations in a single-user single-server MEC network setting and demonstrate that the proposed framework effectively decreases both transmission load and computing cost under various configurations of cache size and tolerable service delay.
\end{abstract}

\section{Introduction}
Recent advancements in smart mobile devices have enabled various emerging applications, such as virtual reality (VR) and augmented reality (AR), which require ultra-high communication and computation capabilities in low latency. To minimize these costs while ensuring a high-quality user experience, the MEC network is a promising solution that can push caching and computing resources to access points, base stations, and even mobile devices at the wireless network edge.

Caching can improve bandwidth utilization by placing frequently accessed content closer to users for future use, which is particularly useful due to the high degree of asynchronous content reuse in mobile traffic. Caching policies can be categorized into two types: \textit{static caching} and \textit{dynamic caching}. Static caching policies are generally based on content popularity distribution and involve cache states that remain unchanged over a relatively long period \cite{sundelay}. In contrast, dynamic caching policies involve content placement updates based on instantaneous user request information, such as the least recently used (LRU) and least frequently used (LFU) policy \cite{10.1145/505696.505701}.

A joint pushing and caching design can improve system performance by proactively transmitting content during low-traffic times to satisfy future user demands. Various joint pushing and caching designs exist that aim to maximize bandwidth utilization \cite{sunpush}, effective throughput \cite{weidelay}, minimize traffic load \cite{codedpush}, or reduce transmit energy consumption \cite{energy1}. However, these policies only consider content delivery and do not account for computation, therefore cannot be directly applied to modern mobile traffic services, such as VR delivery.

To effectively serve mobile traffic, previous designs have considered the joint utilization of cache and computing resources at MEC servers to minimize transmission latency \cite{latency1, latency2} or energy consumption \cite{energy3c}. Some designs also aim to minimize transmission data \cite{vrsun}. However, these designs only consider static caching and do not allow for pushing.

To address the aforementioned issues, we propose a joint computing, pushing, and caching policy optimization approach and validate it in a single-user single-server MEC network. Our approach involves the following steps: (1) We formulate the joint optimization problem as an infinite-horizon discounted Markov decision process, where the aim is to minimize both computation cost and transmission dataload. (2) We use the soft actor-critic (SAC) reinforcement learning (RL) algorithm \cite{haarnoja2018soft} to quickly and stably obtain dynamic computing, pushing, and caching policies. Unlike the classic deep Q-learning algorithm, which requires a Q-network with output nodes for all potential actions, SAC learns Q-functions with few parameters, addressing the curse of dimensionality. We designed an action quantization and correction mechanism to enable SAC, which operates in continuous space, to meet our discrete optimization requirements. (3) We present simulation results with various system parameters to demonstrate the effectiveness of our proposed algorithm.

\begin{figure}[t]
\centering
\includegraphics[width=0.3\textwidth]{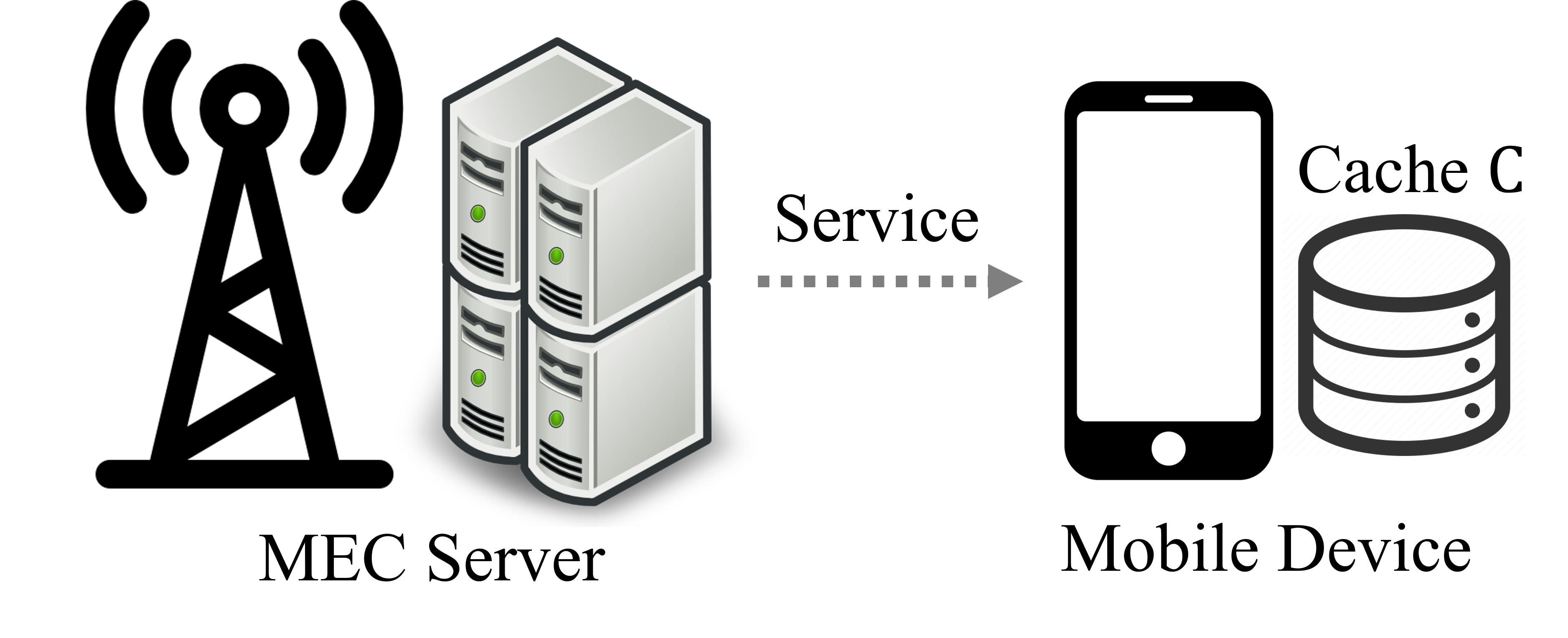}
  \caption{Illustration of MEC network with single MEC server and single mobile device.}
  \label{fig:sys}
  \vspace{-1.5em}
\end{figure}

\section{System Model}
The MEC network we consider comprises a server and a mobile device with caching and computing capabilities, as illustrated in Fig.~\ref{fig:sys}. The MEC server's cache size is large enough to proactively store all input and output data related to tasks requested by the mobile device. In contrast, the mobile device's cache size is limited to $C$ bits. The mobile device is equipped with multi-core computing capabilities, with each core operating at a frequency of $f_D$ cycles per second. We assume that the mobile device has $M$ computing cores. The system operates over an infinite time horizon, and time is slotted in intervals of $\tau$ seconds, with time slots indexed as $t=0,1,2,\cdots$. At the start of each time slot, the mobile device submits a task request that is assumed to be delay-intolerant and must be served before the slot ends.

\subsection{Task Model}
We consider a set of $F$ tasks that can be requested by the mobile device and denote this set as $\mathcal{F} \overset{\Delta}{=} \{1,2, \dots,f,\dots, F\}$. Specifically, each task $f\in \mathcal{F}$ is characterized by a $4$-tuple $\Big\{I_f\ (\text{in\ bits}),\ O_f\  (\text{in\ bits}),\ w_f \ (\text{in cycles/bit}),\ \tau\ (\text{in seconds})\Big\}$. Specifically, $I_f$ represents the size of the remote input data that can be proactively generated from the Internet and cached. The output data size is represented by $O_f$. The parameters $w_f$ and $\tau$ denote the required computation cycles per bit and the maximum allowable service latency.

\subsection{System State}
\subsubsection{Request State}
At the beginning of each time slot $t$, the mobile device generates one task request. Let $A(t) \in \mathcal{F}$ denote the request state of the mobile device, where $A(t)= f$ represents that the mobile device requests task $f$. The cardinality of $\mathcal{F}$ is $F$. Assume that $A(t)$ evolves according to a first-order $F$-state Markov chain, denoted as $\{A(t):t=0,1,2,\dots\}$, which captures both task popularity and inter-task correlation of order one of the task demand process. Let $\Pr[A(t+1)=j | A(t)=i]$ denote the transition probability of going to state $j \in \mathcal{F}$ at time slot $t+1$, given that the request state at time slot $t$ is $i\in \mathcal{F}$ for the task demand process. Assume that $\{A(t)\}$ is time-homogeneous. \textcolor{black}{Denote with $\textbf{Q}\triangleq\big(q_{i,j}\big)_{i\in\mathcal{F},j \in\mathcal{F}}$ the transition probability matrix of $\{A(t)\}$, where $q_{i,j}\triangleq \Pr \left[A(t+1)=j|A(t)=i\right]$.}  
Furthermore, we restrict our attention to irreducible Markov \textcolor{black}{chain} \textcolor{black}{and denote with $\textbf{p} \triangleq (p_{f})_{f\in \mathcal{F}}$ the limiting distribution of $\{A(t)\}$, \textcolor{black}{where $p_{f} \triangleq \lim_{t \to \infty} \Pr[A(t)=f]$. Note that $p_{f} = \sum_{i\in \mathcal{F}} p_{i}q_{i,f}$ for all $f\in \mathcal{F}$}}. 

\subsubsection{Cache State} 
Let $S^I_f(t) \in \{0,1\}$ denote the cache state of the input data for task $f$ in the storage of the mobile device, where $S^I_f(t) = 1$ means that the input data for task $f$ is cached in the mobile device, and $S^I_f(t) = 0$,  otherwise. Let $S^O_f(t) \in \{0,1\}$ denote the cache state of the output data for task $f$ in the storage of the mobile device, where $S^O_f(t) = 1$ means that the output data for task $f$ is cached in the mobile device, and $S^O_f(t) = 0$, otherwise. Denote with $C$ (in bits) the size of the cache space at the mobile device, and the cache size constraint is given by
\begin{equation}
\label{cachesize}
\sum_{f=1}^F I_{f}S^I_{f}(t) + O_fS^O_f(t) \leq C.
\end{equation}

Let $\mathbf{S}(t) \triangleq (S^I_{f}(t), S^O_f(t))_{f\in\mathcal{F}} \in \textcolor{black}{\mathcal{S}}$ denote the cache state of the mobile device at time slot $t$, where \textcolor{black}{$\mathcal{S} \triangleq \{(S^I_{f},S^O_f)_{f\in\mathcal{F}}\in\{0,1\}^F \times \{0,1\}^F: \sum_{f\in \mathcal{F}} I_fS^I_{f}+O_fS^O_f \leq C\}$} represents the cache state space of the mobile device. The cardinality of $\mathcal{S}$ is bounded by $\binom{F}{N_{\min}}$ and $\binom{F}{N_{\max}}$ from below and above, respectively, where $N_{\min} \triangleq  \frac{ C}{\max_{f\in \mathcal{F}} {\{I_f,O_f\}}}$,  and $N_{\max} \triangleq  \frac{ C}{\min_{f\in \mathcal{F}} {\{I_f,O_f\}}}$.

\subsubsection{System State}
At time slot $t$, the system state consists of both system request state and system cache state, denoted as $\textbf{X}(t)\triangleq \left(A(t),\textbf{S}(t)\right)\in$ \textcolor{black}{$\mathcal{F}\times \mathcal{S}$}, where \textcolor{black}{$\mathcal{F}\times \mathcal{S}$} represents the system state space. 
\vspace{-1mm}
\subsection{System Action} \label{sec:sys_action}
\subsubsection{Reactive  Computation Action}
At time slot $t$, we denote with $B^R(t)$ and $E^R(t)$ the reactive transmission bandwidth cost and the reactive computation energy cost. Based on the system state $\textbf{X}(t)=\left(A(t),\textbf{S}(t)\right)$,  the task request $A(t)$ is served as below:
\begin{itemize}
    \item If $S_{A(t)}^O(t) = 1$, the output of task $A(t)$ can be directly obtained from the local cache without any transmission or computation. In this way, the delay is negligible, and the reactive computation energy or transmission cost is zero. 
    
    \item If $S_{A(t)}^I(t)=1$ and $S_{A(t)}^O(t) = 0$, the mobile device can directly compute the task based on the locally cached input data. Let $c_{R,f}(t)\in \left\{1,\cdots,M\right\}$ denote the number of computation cores at the mobile device allocated for reactively processing task $f$ at time slot~$t$. Thus, we directly have $c_{R,f}(t) =0$ for $\forall f\in \mathcal{F}\backslash A(t)$. In order to serve the requested task $A(t)$ within $\tau$, we assume that $\frac{I_{f}w_{f}}{\tau} \textbf{1}(A(t)=f) \leq M f_D, \forall f \in \mathcal{F}$ and require $\frac{I_{A(t)}w_{A(t)}}{\tau} \leq c_{R,A(t)}(t) f_D$. The energy consumed for computing one cycle with frequency $c_{R,f}(t)f_D$ at the mobile device is $\mu c_{R,f}^2(t) f_D^2$, where $\mu$ is the effective switched capacitance related to the chip architecture indicating the power efficiency of CPU at the mobile device.
    The reactive computation energy cost $E^R(t)$ is given by $\mu c_{R,A(t)}^2(t)f_D^2I_{A(t)}w_{A(t)}$, and the reactive transmission cost $B^R(t)$ is  zero.
    
    \item If $S_{A(t)}^I(t)=0$ and $S_{A(t)}^O(t) = 0$, the mobile device should first download the input data of the task $A(t)$ from the MEC server, and then compute it locally. The required latency is given by  $\frac{I_{A(t)}}{B^R(t)}+\frac{I_{A(t)}w_{A(t)}}{c_{R,A(t)}(t)f_D}$.  In order to satisfy the latency constraint, i.e., $\frac{I_{A(t)}}{B^R(t)}+\frac{I_{A(t)}w_{A(t)}}{c_{R,A(t)}(t)f_D} \leq \tau$, the minimum reactive transmission cost $B^R(t)$ is given by $\frac{I_{A(t)}}{\tau - \frac{I_{A(t)}w_{A(t)}}{c_{R,A(t)}(t)f_D}}$, and the reactive computation energy cost $E^R(t)$ is given by $\mu c_{R,A(t)}^2(t) f_D^2I_{A(t)}w_{A(t)}$.
\end{itemize}

In summary, at time slot $t$, the reactive computation action $c_{R,f}(t)$ should satisfy 
\begin{align}\label{reacom}
   c_{R,f}(t) \leq \textbf{1}(A(t) = f)\left(1-S_f^O(t)\right)M, \ \forall f \in \mathcal{F}, 
\end{align}
\noindent and the reactive transmission cost $B^R(t)$ is given by
\begin{equation}\label{reactivebandwidth}
B^R(t) = \left(1-S_{A(t)}^I(t)\right)\left(1-S_{A(t)}^O(t)\right)\frac{I_{A(t)}}{\tau - \frac{I_{A(t)}w_{A(t)}}{c_{R,A(t)}(t)f_D}}, 
\end{equation}
and the reactive computation cost $E^R(t)$ is given by
\begin{equation}\label{reactiveenergy}
    E^R(t) = \left(1-S_{A(t)}^O(t)\right)\mu c_{R,A(t)}^2(t)f_D^2I_{A(t)}w_{A(t)}. 
\end{equation}

Denote with $\textbf{c}_R \triangleq (c_{R,f})_{f\in \mathcal{F}}\in \Pi_C^R(\textbf{X})$ the system reactive computation action, where $\Pi_C^R(\textbf{X}) \triangleq \left\{(c_{R,f})_{f\in \mathcal{F}} \in \left\{0,1,\cdots,M\right\}^F: (\ref{reacom}) \right\}$ denotes the system reactive computation decision space under system state $\textbf{X}$. From (\ref{reacom}), we can see that the cardinality of reactive computation action space is $M+1$.

\subsubsection{Proactive Transmission or Pushing Action} 
Denote with $b_{f}(t) \in \{0,1\}$ the pushing decision of task $f \in \mathcal{F}$, where $b_{f}(t)=1$ means that the remote input data of task $f$ is pushed to the mobile device, and $b_{f}(t) = 0$, otherwise.  Assume that by the end of the time slot, the pushed data are transmitted to the mobile device. In order to satisfy the latency constraint, we have $\frac{\sum_{f=1}^FI_fb_{f}(t)}{\tau} \leq B^P(t)$, where $B^P(t)$ denotes the proactive transmission bandwidth cost. Thus, the minimum proactive transmission cost is given by
\begin{equation}\label{probandwidth}
    B^P(t) = \frac{\sum_{f=1}^F I_fb_{f}(t)}{\tau}.
\end{equation}

In summary, denote with $\mathbf{b} \triangleq \left(b_{f}\right)_{f\in \mathcal{F}} \in \{0,1\}^F$ the system pushing action, where $2^F$ represents the system pushing action space under system state $\textbf{X}$.

\subsubsection{Cache Update Action}
The cache state of each task $f\in \mathcal{F}$ is updated according to 
\begin{align}
    &S_f^I(t+1) = S_f^I(t) + \Delta s_f^I(t),\label{inputupdate}\\
    &S_f^O(t+1) = S_f^O(t) + \Delta s_f^O(t),\label{outputupdate}
\end{align}

\noindent where $\Delta s_f^I(t) \in \{-1,0,1\}$ and $\Delta s_f^O(t) \in \{-1,0,1\}$ denote the update action for the cache state of the input and output data of task $f$, respectively. Then, we have $\forall f \in \mathcal{F}$
\begin{align}
    & -S_f^I(t) \leq \Delta s_f^I(t) \leq \min \left\{b_f(t)+c_{R,f}(t), 1-S_f^I(t)\right\} \label{cachinp1} \\
    & -S_f^O(t) \leq  \Delta s_f^O(t) \leq \min \left\{c_{R,f}(t), 1-S_f^O(t)\right\},
    \label{cachoutp1}\\
    & \resizebox{.88\hsize}{!}{$\sum_{f=1}^F I_f\left(S_f^I(t)+\Delta s_f^I(t)\right)+O_f\left(S_f^O(t)+\Delta s_f^O(t)\right) \leq C$},\label{caupsize}
\end{align}
where the left-hand side of Eq.~(\ref{cachinp1}) indicates that only when the input of task $f$ is cached at the mobile device, can it be removed, the right-hand side of Eq.~(\ref{cachinp1}) indicates that only when the input of task $f$ is proactively transmitted from the MEC server or is reactively transmitted, i.e., $b_f(t) = 1$ or $c_{R,f}(t) > 0$, and has not been cached before, can it be cached into the mobile device. The left-hand side of Eq.~(\ref{cachoutp1}) indicates that only when the output of task $f$ is cached at the mobile device, can it be removed, and the right-hand side of Eq.~(\ref{cachoutp1}) indicates that only when the output of task $f$ is reactively computed at the mobile device, i.e., $c_{R,f}(t) > 0$, and has not been cached before, can it be cached into the mobile device. Eq.~(\ref{caupsize}) indicates that the updated cache state should satisfy the cache size constraint. 

In summary, denote with $\Delta \textbf{s} \triangleq \left(\Delta s_f^I,\Delta s_f^O\right)_{f\in \mathcal{F}} \in \Pi_{\Delta s}(\textbf{X})$ the system cache update action, where $\Pi_{\Delta s}(\textbf{X}) \triangleq \left\{\! \left(\Delta s_f^I,\Delta s_f^O\right)_{f\in \mathcal{F}}\! \in\! \{-1,0,1\}^F\! \times \!\{-1,0,1\}^F: (\ref{cachinp1}),\! (\ref{cachoutp1}),\! (\ref{caupsize})\right\}$ denotes the system cache update action space under system state $\textbf{X}$. 

\subsubsection{System Action}
At each time slot, the system action consists of the reactive computation action, proactive computation action, pushing action, and cache update action, denoted as $\left(\textbf{c}_{R},\textbf{b},\Delta \textbf{s}\right)  \in \Pi(\textbf{X})$, where $\Pi(\textbf{X}) \triangleq \Pi_C^R(\textbf{X})\times \{0,1\}^F\times \Pi_{\Delta s}(\textbf{X})$ denotes the system action space under system state $\textbf{X}$. 
\vspace{-1.8mm}
\subsection{System Cost}
At time slot $t$, the system cost consists of the transmission bandwidth cost and the computation energy cost. In particular, the transmission bandwidth cost consists of both the reactive and proactive transmission costs, given by
\begin{equation}\label{bandwidth}
    B(t) = B^R(t) + B^P(t),
\end{equation}

\noindent where $B^R(t)$ is given in Eq.~(\ref{reactivebandwidth}) and $B^P(t)$ is given in Eq.~(\ref{probandwidth}). The computation energy cost is the reactive computation cost only, i.e.,
\begin{equation}\label{energy}
    E(t) = E^R(t),
\end{equation}
\noindent where $E^R(t)$ is given in Eq.~(\ref{reactiveenergy}). To balance the communication and computation cost, we choose the weighted sum $B(t)+\lambda E(t)$ as the system cost at time slot $t$. 
\vspace{-2mm}
\section{Problem Formulation}
Given an observed system state $\textbf{X}$, the joint reactive computing, transmission, and caching action, denoted as $\left(\textbf{c}_{R}, \textbf{b},\Delta \textbf{s}\right)$, is determined according to a policy defined as below. 
\begin{Def}[Stationary Joint Computing, Pushing and Caching Policy] A stationary joint computing, pushing, and caching policy $\pi$ is a mapping from system state $\textbf{X}$ to system action $\left(\textbf{c}_{R},\textbf{b},\Delta \textbf{s}\right)$, i.e., $\left(\textbf{c}_{R},\textbf{b},\Delta \textbf{s}\right)= \pi(\textbf{X}) \in \Pi(\textbf{X})$.
\end{Def}

From properties of $\{A(t)\}$ and $\{\textbf{S}(t)\}$, the induced system state process $\{\textbf{X}(t)\}$ under policy $\pi$ is a controlled Markov chain. The expected total discounted cost is given as
\begin{align}
    \phi(\pi)\triangleq \limsup_{T\rightarrow \infty} \sum_{t=0}^{T-1}\gamma^{t}\mathbb{E}\left[B(t)+\lambda E(t)\right],
\end{align}
where the expectation is taken over the task request process. 

In this paper, we aim to obtain optimal joint computing, pushing, and caching policy to minimize the sum of infinite horizon discounted system cost, i.e., minimize both the transmission and computation cost. 

\begin{Prob}[Joint Computing, Pushing and Caching Policy Optimization]
    \begin{align}
        \phi^* \triangleq & \min_{\pi} \ \ \phi(\pi) \nonumber\\
        &\  s.t.\ \  \pi(\textbf{X}) \in \Pi(\textbf{X}),\ \  \forall   \textbf{X} \in \mathcal{F}\times \mathcal{S}. \nonumber
    \end{align}
\end{Prob}

\section{Soft Actor-Critic Learning}

\subsection{SAC System State and Action} \label{sec:SAC_action}

The system state $\mathbf{x}$ of SAC is designed to match the system state $\textbf{X}$ in the formulated problem, such that $\mathbf{x}=\textbf{X}=\left(A(t),\textbf{S}(t)\right)$, with a vector size of $2F+1$. 

The SAC algorithm is designed to solve continuous-action problems, whereas the required system action $\left(\textbf{c}_{R},\textbf{b},\Delta \textbf{s}\right)$ in the formulated problem is discrete. To address this issue, we define the system action of the SAC as the \textbf{\textit{continuous version}} of the formulated system action space. This continuous version is denoted as $\mathbf{a}=\left(\bar{\textbf{c}}_{R},\bar{\textbf{b}},\Delta \bar{\textbf{s}}\right) \in \bar{\Pi}(\textbf{X}) \triangleq \bar{\Pi}_C^R(\textbf{X})\times [0,1]^F\times \bar{\Pi}_{\Delta s}(\textbf{X})$. 

As $\bar{\textbf{c}}_{R}\triangleq \left\{(\bar{c}_{R,f})_{f\in \mathcal{F}}\right\}$ must always equal zero for $f\in \mathcal{F}{\setminus} A(t)$, the action space of SAC can be simplified by disregarding the computing cores for non-requested tasks. We can obtain the simplified form of action $\mathbf{a}$ as $\mathbf{a}=\left(\bar{c}_{A(t)},\bar{\textbf{b}},\Delta \bar{\textbf{s}}\right)$, with a vector size of $3F+1$.
\vspace{-1mm}
\subsection{SAC Learning}
\label{sec:sac_learning}
SAC is an off-policy deep reinforcement learning method that maintains the advantages of entropy maximization and stability while offering sample-efficient learning \cite{haarnoja2018soft}. It operates on an actor-critic framework where the actor is responsible for maximizing expected reward while simultaneously maximizing entropy. The critic evaluates the effectiveness of the policy being followed.

A general form of maximum-entropy RL is given by
\begin{equation}
    J(\pi)=\sum_{t=0}^T \mathbb{E}_{\left(\mathbf{x}_t, \mathbf{a}_t\right) \sim \rho_\pi}\left[r\left(\mathbf{x}_t, \mathbf{a}_t\right)+\alpha \mathcal{H}\left(\pi\left(\cdot \mid \mathbf{x}_t\right)\right)\right],
    \label{eq:ge_rl}
\end{equation}
where the temperature parameter $\alpha$ determines the relative importance of the entropy term against the reward $r$, and the entropy term is given by $\mathcal{H}\left(\pi\left(\cdot \mid \mathbf{x}_t\right)\right)=\mathbb{E}_{\mathbf{a}_t}\left[-\log \pi\left(\mathbf{a}_t \mid \mathbf{x}_t\right)\right]$.

The SAC algorithm \cite{haarnoja2018soft} is a policy iteration approach designed to solve the optimization problem in Eq.~\eqref{eq:ge_rl}. It comprises two essential components: soft Q-function $Q_{\theta}\left(\mathbf{x}_{t}, \mathbf{a}_{t}\right)$, and policy $\pi_{\phi}\left(\mathbf{a}_{t} \mid \mathbf{x}_{t}\right)$. To deal with the large continuous domains, neural networks approximate these components, with the network parameters denoted by $\theta$ and $\phi$. For example, the policy is modeled as a Gaussian distribution with a fully connected network providing the mean and covariance value, and the Q-function is also approximated using a fully connected neural network. Following \cite{haarnoja2018soft}, the update rules for $\theta$ and $\phi$ are provided below.

The soft Q-function parameters can be trained to minimize the soft Bellman residual
\begin{equation}
    \label{eq:J_Q}
    \begin{aligned} 
    J_{Q}(\theta)=\mathbb{E}_{\left(\mathbf{x}_{t}, \mathbf{a}_{t}\right) \sim \mathcal{D}}\Bigl[ 
    & \frac{1}{2}\bigl(Q_{\theta}\left(\mathbf{x}_{t}, \mathbf{a}_{t}\right)-\bigl(r\left(\mathbf{x}_{t}, \mathbf{a}_{t}\right)+ \\
    & \gamma \mathbb{E}_{\mathbf{x}_{t+1} \sim p}\left[V_{\bar{\theta}}\left(\mathbf{x}_{t+1}\right)\right]\bigr)\bigr)^{2}\Bigr],
    \end{aligned}
\end{equation}
where $\mathcal{D}$ is the distribution of previously sampled states and actions, $p$ is the transition probability between states, and the value function $V_{\bar{\theta}}(\mathbf{x}_t)$ is implicitly parameterized through the soft Q-function parameters as follows:
\begin{equation}
V_{\bar{\theta}}\left(\mathbf{x}_t\right)=\mathbb{E}_{\mathbf{a}_t \sim \pi}\left[Q_{\bar{\theta}}\left(\mathbf{x}_t, \mathbf{a}_t\right)-\alpha \log \pi\left(\mathbf{a}_t \mid \mathbf{x}_t\right)\right].
\end{equation}

The update makes use of a target soft Q-function $Q_{\bar{\theta}}$ with parameters $\bar{\theta}$ obtained as an exponentially moving average of the soft Q-function weights $\theta$, which helps stabilize training. The soft Bellman residual $J_{Q}(\theta)$ in Eq.~\eqref{eq:J_Q} can be optimized with stochastic gradients
\begin{equation}
    \label{eq:J_Q_grad}
    \begin{aligned} 
   \hat{\nabla}_\theta J_Q(\theta)= & \nabla_\theta Q_\theta\left(\mathbf{a}_t, \mathbf{x}_t\right)\Bigl(Q_\theta\left(\mathbf{x}_t, \mathbf{a}_t\right)-\bigl(r\left(\mathbf{x}_t, \mathbf{a}_t\right) + \\ &
  \gamma\left(Q_{\bar{\theta}}\left(\mathbf{x}_{t+1}, \mathbf{a}_{t+1}\right)-\alpha \log \left(\pi_\phi\left(\mathbf{a}_{t+1} \mid \mathbf{x}_{t+1}\right)\right)\right)\bigr)\Bigr).
    \end{aligned}
\end{equation}

The policy parameters $\phi$ can be learned by directly minimizing the expected KL divergence in
\begin{equation}
    \label{eq:J_Pi}
    \begin{aligned} 
    J_{\pi}(\phi)=\mathbb{E}_{\mathbf{x}_{t} \sim\mathcal{D}}\Bigl[\mathbb{E}_{\mathbf{a}_{t} \sim \pi_{\phi}}\bigl[
    & \alpha \log \left(\pi_{\phi}\left(\mathbf{a}_{t} \mid \mathbf{x}_{t}\right)\right)- \\
    & Q_{\theta}\left(\mathbf{x}_{t}, \mathbf{a}_{t}\right)\bigr]\Bigr].
    \end{aligned}
\end{equation}

A neural network transformation is used to parameterize the policy as $\mathbf{a}_{t}=f_{\phi}\left(\epsilon_{t} ; \mathbf{x}_{t}\right)$, where $\epsilon_{t}$ is an input noise vector sampled from a Gaussian distribution. The objective stated by Eq.~\eqref{eq:J_Pi} can be rewritten as
\begin{equation}
    \label{eq:J_Pi_v2}
    \begin{aligned} 
    J_{\pi}(\phi)=\mathbb{E}_{\mathbf{x}_{t} \sim \mathcal{D}, \epsilon_{t} \sim \mathcal{N}}\bigl[
    & \alpha \log \pi_{\phi}\left(f_{\phi}\left(\epsilon_{t} ; \mathbf{x}_{t}\right) \mid \mathbf{x}_{t}\right)
    \\
    & -Q_{\theta}\left(\mathbf{x}_{t}, f_{\phi}\left(\epsilon_{t} ; \mathbf{x}_{t}\right)\right)\bigr],
    \end{aligned}
\end{equation}
where $\pi_{\phi}$ is defined implicitly in terms of $f_{\phi}$. The gradient of Eq.~\eqref{eq:J_Pi_v2} is approximated with
\begin{equation}
    \label{eq:J_Pi_grad}
    \begin{aligned} 
    \hat{\nabla}_{\phi} J_{\pi}(\phi)=
    & \nabla_{\phi} \alpha \log \left(\pi_{\phi}\left(\mathbf{a}_{t} \mid \mathbf{x}_{t}\right)\right)+ 
    \bigl(\nabla_{\mathbf{a}_{t}} \alpha \log \left(\pi_{\phi}\left(\mathbf{a}_{t} \mid \mathbf{x}_{t}\right)\right) \\
    & -\nabla_{\mathbf{a}_{t}} Q\left(\mathbf{x}_{t}, \mathbf{a}_{t}\right)\bigr) \nabla_{\phi} f_{\phi}\left(\epsilon_{t} ; \mathbf{x}_{t}\right),
    \end{aligned}
\end{equation}
where $\mathbf{a}_{t}$ is evaluated using $f_{\phi}\left(\epsilon_{t} ; \mathbf{x}_{t}\right)$. 

\noindent \textbf{Remark}: In the maximum entropy framework, the soft policy iteration that alternates between the policy evaluation Eq.~\eqref{eq:J_Q} and the policy improvement Eq.~\eqref{eq:J_Pi} converges to the optimal policy. \textit{Proof in} \cite{haarnoja2018soft}.

\subsection{Action Quantization and Correction}
The SAC learning algorithm produces the SAC action $\mathbf{a}_t$ at time $t$ that maximizes the policy value $\pi_{\phi}\left(\mathbf{a}_{t} \mid \mathbf{x}_{t}\right)$ given the SAC state $\mathbf{x}_t$. However, to evaluate the reward and update the cache, we need to convert the continuous SAC action $\mathbf{a}_t$ into a discrete action $\left(c_{A(t)},\textbf{b},\Delta\textbf{s}\right)$. We achieve this through a simple action quantization method that involves thresholding and integer projection.

\noindent \textbf{Action quantization}: Consider an element $\bar{\eta}$ in the SAC action $\mathbf{a}$ and its corresponding quantized version $\eta$ in the system action, where $\eta$ belongs to the selection set $S_{\eta}$. To convert $\bar{\eta}$ to $\eta$, we employ uniform thresholding for integer projection, which is given by
\begin{equation}
\label{eq:action_quan}
\eta = \min{S_{\eta}} + (\bar{\eta}-\min{S_{\eta}}) \, \text{mod} \, \frac{\max{S_{\eta}}-\min{S_{\eta}}}{\max{S_{\eta}}-\min{S_{\eta}} + 1}
\end{equation}

As an example, consider the push action $b_f(t)\in S_{b_f}=\{0,1\}$, we can determine its quantized value $b_f(t) = \bar{b}_f (t) \mod 0.5$ using Eq.~\eqref{eq:action_quan}.
\vspace{-1mm}

\noindent \textbf{Action correction}: Due to the constraints outlined in Eqs.~\eqref{reacom}, \eqref{reactivebandwidth}, \eqref{cachinp1}, \eqref{cachoutp1}, and \eqref{caupsize}, the valid action space of the system is very limited and sparsely spanned, with a cardinality of $(M+1)\times 2^F \times 3^{2F}$. Consequently, even with techniques such as penalty reward, it is difficult for the SAC algorithm to learn which actions are valid in the huge and sparsely spanning action space. Therefore, the post-quantization action $\left(c_{A(t)},\textbf{b},\Delta\textbf{s}\right)$ is usually not valid. To overcome this issue, we propose to correct the output action of SAC and make it valid using \textit{Rules 1, 5, and 7} as detailed below. These rules are designed to satisfy the constraints presented in Section~\ref{sec:sys_action}. Additionally, we suggest some general \textit{Rules 2, 3, 4, and 6} to accelerate training and refine system action.
\begin{itemize}
\item \textit{Rule 1}: when $S^O_{A(t)} {=} 0$, if the $c_{A(t)}$ is smaller than the minimum workable value $	\lceil I_{A(t)} w_{A(t)} / (\tau f_D) \rceil$, we will correct it by $c_{A(t)}=\lceil I_{A(t)} w_{A(t)} / (\tau f_D) \rceil$; when $S^O_{A(t)} = 0$,  $c_{A(t)}=0$. This is to constrain the total latency.

\item \textit{Rule 2}: when $S^I_f + S^O_f \ge 1$, $b_f=0$. It indicates no need for proactive pushing if any data of a task is cached.

\item \textit{Rule 3}: there is at most one task being proactively transmitted for the task with largest $\bar{b}_f$ and $b_f=1$. Other tasks are corrected to $b_f=0$. This is to reduce the unnecessary pushing cost given that the mobile device has one task request at each time slot.

\item \textit{Rule 4}: if $b_f=1$, $\Delta s^I_f=1$. It indicates that the proactive pushing data has to be cached.

\item \textit{Rule 5}: if the cache sum Eq.~\eqref{caupsize} exceeds the capacity, we drop the input or output cache according to the ascending order of the $\bar{\textbf{s}}$ values until the capacity fits.

\item \textit{Rule 6}: if the cache sum Eq.~\eqref{caupsize} is inferior to capacity, we try to add the reactive input or output cache according to the descending order of $\Delta \bar{s}^I_{A(t)}$ and $\Delta \bar{s}^O_{A(t)}$ values.

\item\textit{Rule 7}: clip the cache action $\Delta \textbf{s}$ according to the $\min, \max$ limit in Eq.~\eqref{cachinp1} Eq.~\eqref{cachoutp1}.
\end{itemize}

\begin{algorithm}[t]
\caption{SAC Learning for Our Problem }\label{alg:alg1}
\begin{algorithmic}
\STATE 
\STATE Initialize parameters $\theta, \bar\theta, \phi$ for networks $Q_{\theta}$, $Q_{\bar\theta}$, $\pi_{\phi}$.
\STATE Initialize learning rate $\lambda_Q$, $\lambda_\pi$, and weight $\xi$.
\STATE \textbf{for} each iteration \textbf{do}
\STATE \hspace{0.5cm} \textbf{for} each environment step \textbf{do}
\STATE \hspace{1.0cm} $\mathbf{a}_t \sim \pi_\phi\left(\mathbf{a}_t \mid \mathbf{x}_t\right)$
\STATE \hspace{1.0cm} $\mathbf{x}_{t+1} \sim p\left(\mathbf{x}_{t+1} \mid \mathbf{x}_t, \mathbf{a}_t\right)$
\STATE \hspace{1.0cm} $\mathbf{a}_t$ quantization \& correction, calculate $r\left(\mathbf{x}_t, \mathbf{a}_t\right)$
\STATE \hspace{1.0cm} $\mathcal{D} \leftarrow \mathcal{D} \cup\left\{\left(\mathbf{x}_t, \mathbf{a}_t, r\left(\mathbf{x}_t, \mathbf{a}_t\right), \mathbf{x}_{t+1}\right)\right\}$
\STATE \hspace{0.5cm} \textbf{end for}

\STATE \hspace{0.5cm} \textbf{for} each gradient step \textbf{do}
\STATE \hspace{1.0cm} $\theta_i \leftarrow \theta_i-\lambda_Q \hat{\nabla}_{\theta_i} J_Q\left(\theta_i\right) \text { for } i \in\{1,2\}$
\STATE \hspace{1.0cm} $\phi \leftarrow \phi-\lambda_\pi \hat{\nabla}_\phi J_\pi(\phi)$
\STATE \hspace{1.0cm} $\bar \theta_i \leftarrow \xi \theta_i + (1-\xi) \bar \theta_i \text { for } i \in\{1,2\}$
\STATE \hspace{0.5cm} \textbf{end for}
\STATE \textbf{end for}

\end{algorithmic}
\end{algorithm}
\vspace{-2mm}
\subsection{Reward Design}
The reward $r(\mathbf{x}, \mathbf{a})$ of SAC state $\mathbf{x}$ and action $\mathbf{a}$ are designed to be a function of resulting bandwidth and computation cost 
\begin{equation}
r(\mathbf{x}, \mathbf{a}) = - \kappa (B(t)+\lambda E(t))
\end{equation}
\noindent where $\kappa$ is the normalization coefficient, and is set as $10^{-6}$.

The complete SAC learning algorithm is presented in Algorithm~\ref{alg:alg1}, where $\lambda_Q, \lambda_\pi$ are the step sizes (or learning rate) for stochastic gradient descent, and are chosen to be $1\times10^{-4}$. $\xi$ is the target smoothing coefficient chosen to be 0.005.

\section{Implementation}
We generated simulated data for training and testing by randomly generating a Markov chain from the task set $\mathcal{F}$. The transition probability of a task $i$ to one randomly selected task $j \in \mathcal{F}\backslash i$ was set to the maximum transition probability, i.e., $p_{i, j} = p_{\max}$. The probability to other tasks $k \in \mathcal{F}\backslash j$ was given by $p_{i, k} = (1-p_{i, j})\frac{|p^\prime_{i, k}|}{\sum_{f \in \mathcal{F}\backslash j}|p^\prime_{i, f}|}$, where $p^\prime_{i, k}$ or $p^\prime_{i, f}$ were random samples from a uniform distribution. This designed Markov chain represents the request popularity and transition preference of $F$ tasks. We sampled $10^6$ requested tasks using a frame-by-frame approach. In our simulation, we considered $M=8$, $F=4$, a maximum transition probability of $0.7$, $\lambda=1$, $I_f = 16000$ bits, $O_f = 30000$ bits, $w=800$ cycles/bit, $\tau=0.02$ seconds, $f_D = 1.7\times 10^8$ cycles/s, $\mu= 10^{-19}$, and $C=40\times 10^3$ bits.

For ease of training and stabilization, both the SAC action $\mathbf{a}_t$ and system state $\mathbf{x}_t$ are normalized to the range of $[-1, 1]$. The implementation of the system is done with Python and PyTorch. The training and testing were deployed on a PC with TITAN RTX GPU using batch size 256, discount factor $\gamma=0.99$, automatic entropy temperature $\alpha$ tuning \cite{haarnoja2018soft}, network hidden-layer size 256, one model update per step, one target update per 1000 steps, and replay buffer size of $1\times10^7$. We make 10 testing epochs after every 10 training epochs and stop the training and testing when the reward and loss converge. 


\begin{figure}
\centering
\includegraphics[width=0.4\textwidth]{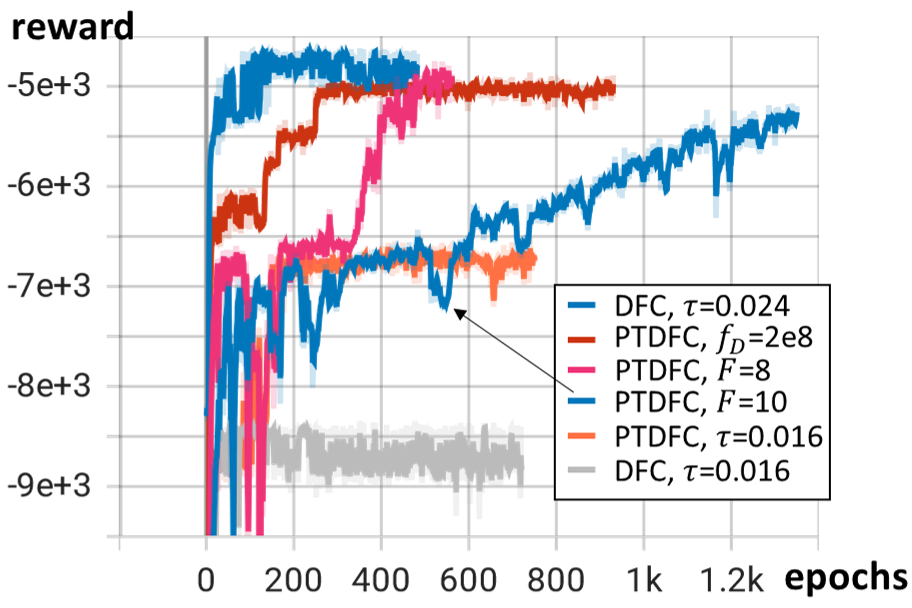}
  \caption{Training convergence of SAC algorithm for different configuration cases.}
  \label{fig:converge}
  \vspace{-1.5em}
\end{figure}

\section{Evaluation and Analysis}
\subsection{Baselines}
The proposed system is built on the \textit{proactive transmission and dynamic-computing-frequency reactive service with cache} \textbf{(PTDFC)}. For comparison, we consider the following baselines:
\begin{itemize}
      \item \textbf{\textit{Most-recently-used proactive transmission and least-recently-used cache replacement} (MRU-LRU)}: A heuristic algorithm \cite{sunpush} that serves the requested task reactively while proactively caching the \textit{input data} of the most-recently-used task and replacing the least-recently-used task's cache when the cache is full. The number of computing cores used is fixed at $0.75M$.
    \item \textbf{\textit{Most-frequently-used proactive transmission and least-frequently-used cache replacement} (MFU-LFU)}: Similar to MRU-LRU, this algorithm replaces the most/least recently used task with the most/least frequently used task.
    \item \textbf{\textit{Dynamic-computing-frequency reactive service with no cache} (DFNC)}: This algorithm reactively serves the requested task at each time slot $t$ by downloading the input data from the MEC server and computing the output data. 
    \item \textbf{\textit{Dynamic-computing-frequency reactive service with cache} (DFC)}: Similar to DFNC, this algorithm also reactively serves the requested task at each time slot $t$ but can cache the input or output data into limited capacity.
\end{itemize}


\subsection{Convergence}
We show the convergence of the SAC algorithm in Fig.~\ref{fig:converge} by plotting the reward vs. epochs curves for different system setups. For setups with a smaller action space, such as those with smaller $\tau$ and no proactive transmission, the SAC algorithm converges quickly in around 200 epochs. However, more complex setups with larger action spaces, such as those with proactive transmission and more tasks $F$, typically take 500 epochs or longer to converge.

\subsection{Different Cache Size $C$}
\begin{figure}[!t]
\centering
\includegraphics[width=0.45\textwidth, trim=0 10 0 30,clip]{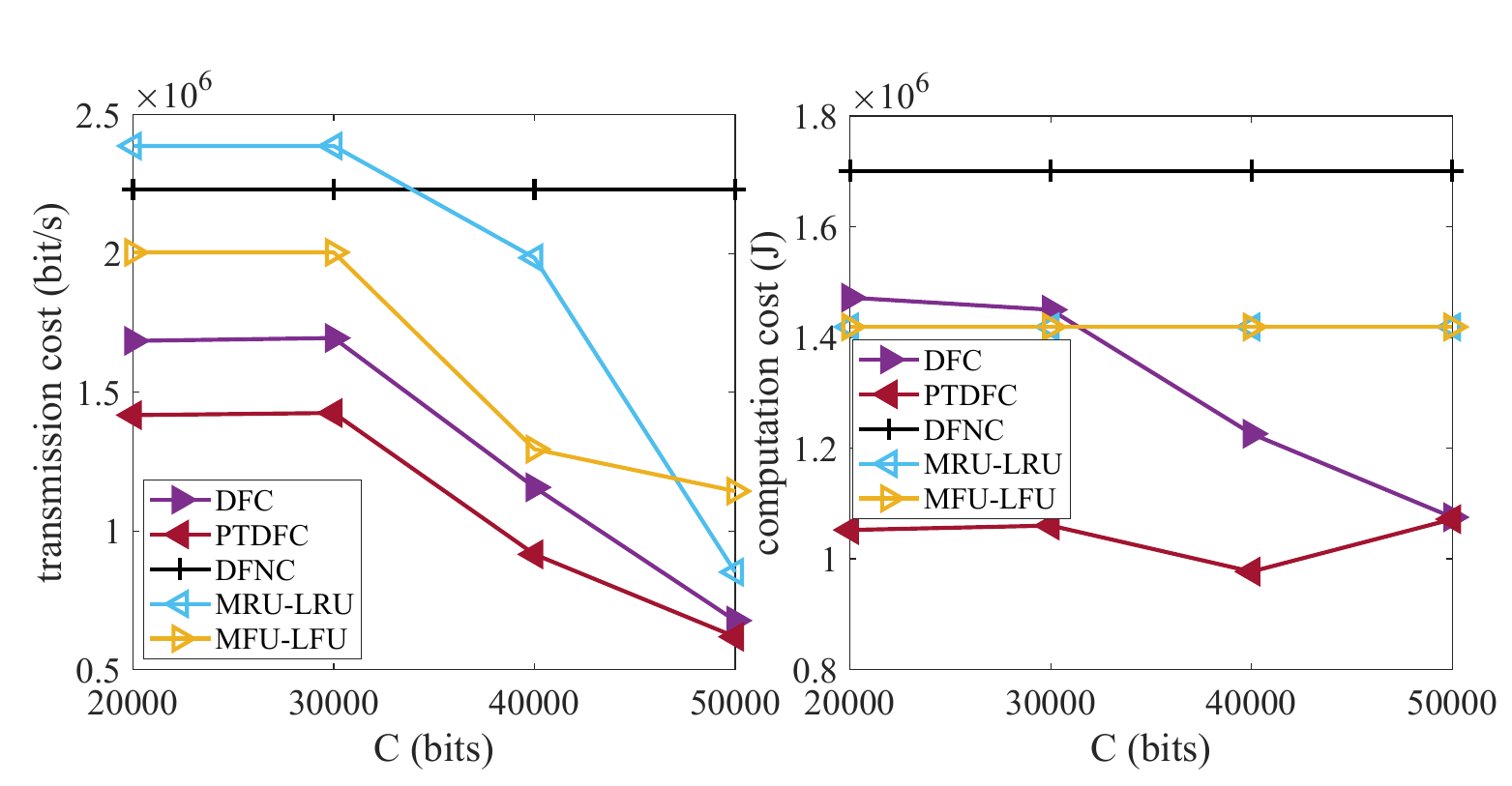}
  \caption{Varying cache size $C$ under default configuration.}
  \label{fig:varyC}
  \vspace{-1.5em}
\end{figure}

The average transmission cost and computation cost of three SAC-enabled algorithms, DFNC, DFC, PTDFC, and two heuristic algorithms MRU-LRU, MFU-LFU, were compared under different cache sizes $C$, as shown in Fig.~\ref{fig:varyC}. While the cache size change did not affect the DFNC algorithm, the rest of the algorithms showed decreasing transmission costs with increasing $C$ due to the ability to cache more input data locally. In addition, the proposed PTDFC algorithm consistently achieved lower transmission and computation costs than the other algorithms under different cache sizes. For very large cache sizes, such as $C=50000$ bits, the performance of PTDFC and DFC was similar.

\subsection{Different Tolerable Service Delays $\tau$}
\begin{figure}[!t]
\centering
\includegraphics[width=0.45\textwidth, trim=0 10 0 30,clip]{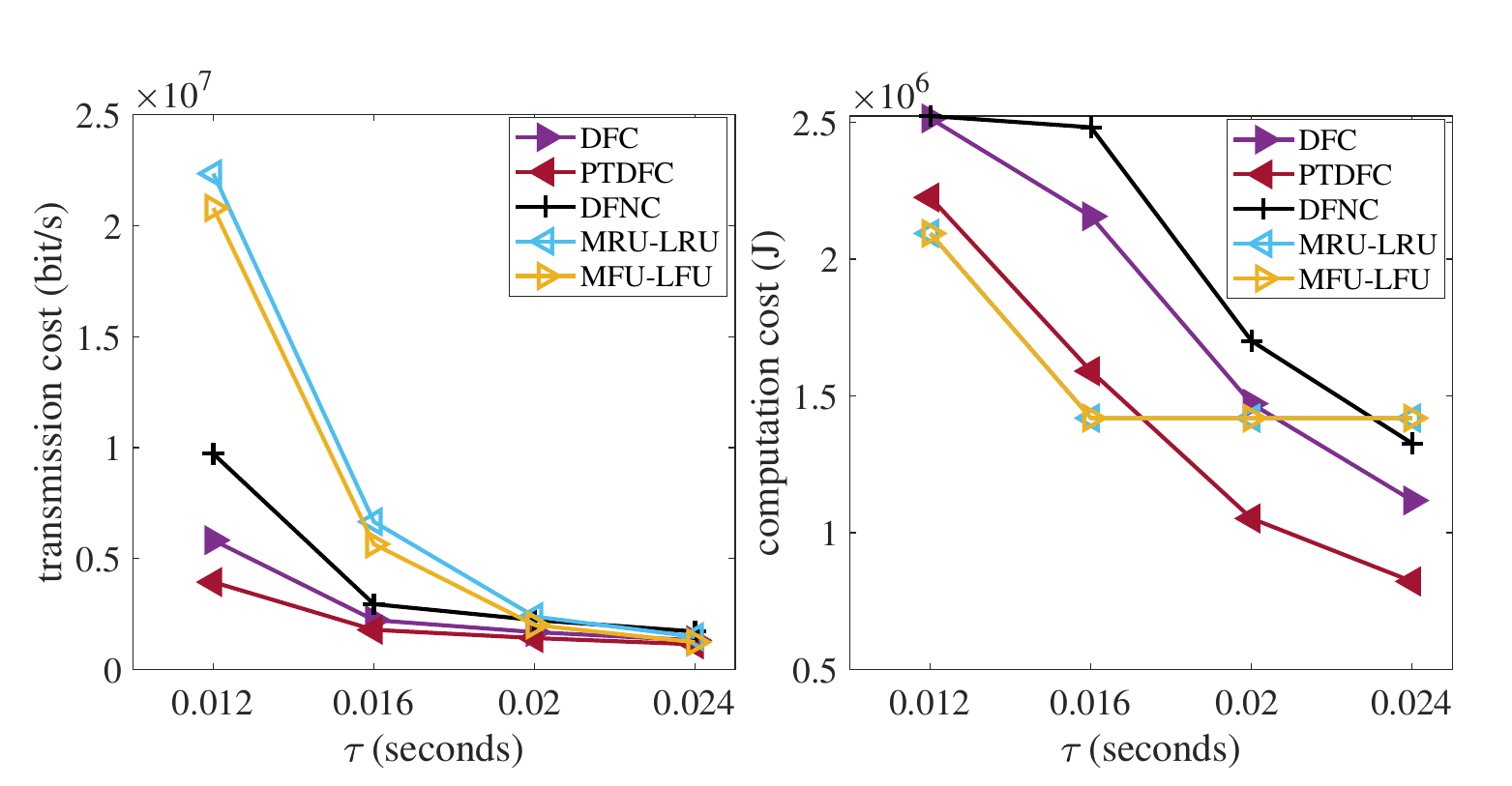}
  \caption{Varying maximum tolerable service latency $\tau$ under default configuration.}
  \label{fig:varyTau}
  \vspace{-1.5em}
\end{figure}

Fig.~\ref{fig:varyTau} illustrates the performance of five algorithms at different maximum tolerable service delays. As $\tau$ increases, the cost of all algorithms decreases because there is more time available for the transmission and computation process, requiring less bandwidth and lower computing frequency. Among the three SAC-enabled algorithms, the proposed PTDFC algorithm consistently achieves the lowest transmission and computation cost under different $\tau$ values. However, as $\tau$ gets larger, the transmission cost of all five algorithms begins to converge, and the advantages of PTDFC become less significant. This is because enough time is available for transmission even with the lowest-frequency reactive computing.


\section{Conclusion}
This paper investigates joint computing, pushing, and caching optimization in a single-user single-server MEC network to reduce the transmission data load and computation cost. A framework based on SAC learning with action quantization and correction techniques is proposed to enable dynamic orchestration of the three activities. Simulation results demonstrate the effectiveness of the proposed framework in reducing both transmission load and computing cost, outperforming baseline algorithms under various parameter settings.
\vspace{-5mm}
\bibliographystyle{IEEEtran}
\bibliography{bibtex}

\begin{thebibliography}{10}
\providecommand{\url}[1]{#1}
\csname url@samestyle\endcsname
\providecommand{\newblock}{\relax}
\providecommand{\bibinfo}[2]{#2}
\providecommand{\BIBentrySTDinterwordspacing}{\spaceskip=0pt\relax}
\providecommand{\BIBentryALTinterwordstretchfactor}{4}
\providecommand{\BIBentryALTinterwordspacing}{\spaceskip=\fontdimen2\font plus
\BIBentryALTinterwordstretchfactor\fontdimen3\font minus
  \fontdimen4\font\relax}
\providecommand{\BIBforeignlanguage}[2]{{%
\expandafter\ifx\csname l@#1\endcsname\relax
\typeout{** WARNING: IEEEtran.bst: No hyphenation pattern has been}%
\typeout{** loaded for the language `#1'. Using the pattern for}%
\typeout{** the default language instead.}%
\else
\language=\csname l@#1\endcsname
\fi
#2}}
\providecommand{\BIBdecl}{\relax}
\BIBdecl

\bibitem{sundelay}
Y.~Sun, Z.~Chen, and H.~Liu, ``Delay analysis and optimization in cache-enabled
  multi-cell cooperative networks,'' in \emph{IEEE Global Communications
  Conference (GLOBECOM)}, 2016, pp. 1--7.

\bibitem{10.1145/505696.505701}
\BIBentryALTinterwordspacing
J.~Wang, ``A survey of web caching schemes for the internet,'' \emph{SIGCOMM
  Comput. Commun. Rev.}, vol.~29, no.~5, p. 36–46, oct 1999. [Online].
  Available: \url{https://doi.org/10.1145/505696.505701}
\BIBentrySTDinterwordspacing

\bibitem{sunpush}
Y.~Sun, Y.~Cui, and H.~Liu, ``Joint pushing and caching for bandwidth
  utilization maximization in wireless networks,'' \emph{IEEE Transactions on
  Communications}, vol.~67, no.~1, pp. 391--404, 2019.

\bibitem{weidelay}
W.~Chen and H.~V. Poor, ``Content pushing with request delay information,''
  \emph{IEEE Transactions on Communications}, vol.~65, no.~3, pp. 1146--1161,
  2017.

\bibitem{codedpush}
Y.~Lu, W.~Chen, and H.~V. Poor, ``Coded joint pushing and caching with
  asynchronous user requests,'' \emph{IEEE Journal on Selected Areas in
  Communications}, vol.~36, no.~8, pp. 1843--1856, 2018.

\bibitem{energy1}
M.~Gregori, J.~G{\'o}mez-Vilardeb{\'o}, J.~Matamoros, and D.~G{\"u}nd{\"u}z,
  ``Wireless content caching for small cell and d2d networks,'' \emph{IEEE
  Journal on Selected Areas in Communications}, vol.~34, no.~5, pp. 1222--1234,
  2016.

\bibitem{latency1}
X.~Yang, Z.~Fei, J.~Zheng, N.~Zhang, and A.~Anpalagan, ``Joint multi-user
  computation offloading and data caching for hybrid mobile cloud/edge
  computing,'' \emph{IEEE Transactions on Vehicular Technology}, vol.~68,
  no.~11, pp. 11\,018--11\,030, 2019.

\bibitem{latency2}
M.~Chen, Y.~Hao, L.~Hu, M.~S. Hossain, and A.~Ghoneim, ``Edge-cocaco: Toward
  joint optimization of computation, caching, and communication on edge
  cloud,'' \emph{IEEE Wireless Communications}, vol.~25, no.~3, pp. 21--27,
  2018.

\bibitem{energy3c}
Y.~Hao, M.~Chen, L.~Hu, M.~S. Hossain, and A.~Ghoneim, ``Energy efficient task
  caching and offloading for mobile edge computing,'' \emph{IEEE Access},
  vol.~6, pp. 11\,365--11\,373, 2018.

\bibitem{vrsun}
Y.~{Sun}, Z.~{Chen}, M.~{Tao}, and H.~{Liu}, ``Communications, caching, and
  computing for mobile virtual reality: Modeling and tradeoff,'' \emph{IEEE
  Trans. Commun.}, vol.~67, no.~11, pp. 7573--7586, Nov. 2019.

\bibitem{haarnoja2018soft}
T.~Haarnoja, A.~Zhou, P.~Abbeel, and S.~Levine, ``Soft actor-critic: Off-policy
  maximum entropy deep reinforcement learning with a stochastic actor,'' in
  \emph{International conference on machine learning}.\hskip 1em plus 0.5em
  minus 0.4em\relax PMLR, 2018, pp. 1861--1870.

\end{thebibliography}

\end{document}